\documentclass[UTF8,a4paper]{article}
\usepackage{amsmath}
\usepackage{graphicx}
\usepackage{subfigure}
\usepackage{epstopdf}
\usepackage{geometry}
\usepackage{ulem}
\usepackage{indentfirst}
\usepackage{amsfonts,amssymb,dsfont}
\usepackage{setspace}
\usepackage{threeparttable}
\usepackage{amsmath,mathtools,amsthm}
\usepackage{array}
\usepackage{extarrows}
\usepackage{upgreek}

\makeatletter
\renewcommand*\env@matrix[1][\arraystretch]{%
  \edef\arraystretch{#1}%
  \hskip -\arraycolsep
  \let\@ifnextchar\new@ifnextchar
  \array{*\c@MaxMatrixCols c}}
\makeatother
\begin{document}


\title{\bf Attractive polaron in a Dirac system within ladder approximation}
\author{Chen-Huan Wu
\thanks{chenhuanwu1@gmail.com}
\\College of Physics and Electronic Engineering, Northwest Normal University, Lanzhou 730070, China}

\maketitle
\vspace{-30pt}
\begin{abstract}
\begin{large}

We investigate the properties of the attractive polaron formed by a single impurity dressed with the particle-hole excitations
in a Dirac system
at zero-temperature limit.
Base on the single particle-hole variational ansatz,
we deduce the expressions of the pair propagator, self-energy (negative), and the non-self-consistent medium $T$-matrix.
Different to the self-consistent $T$-matrix which contains two channels
due to the many-body effect,
the non-self-consistent $T$-matrix discussed in this paper 
contains only the closed channel (i.e., the bare one)
since we consider only a finite number of the majority particles.
The chiral factor is also discussed within the pair propagator due to the scattering feature of the Dirac system
(with a certain scattering angle $a$),
and we found that the fluctuation of the pair propagator is proportional to the term ${\rm cos}\ a$.
Finally, we also discuss the particle spectral function of the polaron in the presence of quantum many-body effect
as well as the pair-propagator at finite temperature.\\
\\
 $PACS\ number(s)$: 71.10.Hf, 71.10.Li, 71.36.+c\\
$ Keywords$: attractive polaron;
non-self consistent medium T-matrix;
self-energy;
pair propagator;
ladder approximation\\

\end{large}

\end{abstract}
\begin{large}

\section{Introduction}

Considering the disorder effect origin from the impurities (in coherent background),
the polaron as an excited quasiparticle in the population/spin-imbalanced Fermi gases or BEC and even the topological insulator
\cite{amacho-Guardian A,Shvonski A,Qin F},
are important when taking the many-body effect into account.
The disorder-induced self-energy\cite{ele}
describes the impurity-Fermions (for Fermionic polaron) or impurity-Bosons (for Bosonic polaron) interaction 
where the impurity dressed by the corresponding particle-hole excitations.
Besides, for Dirac systems, since the spin rotation is missing 
in the presence of Dirac $\delta$-type impurity field,
the spin structure is fixed,
and the interacting spins of impurity and majority particles are usually opposite (due to the Pauli principle)
which provides the opportunity to form the Cooper pair and the strongly bound dimer in superfluid.
That also implies, the low-temperature which with weaker spin relaxation is beneficial 
to the formation of polarons.
We present in Appendix.A a brief discussion of the variational approach in mean-field approximation
which is valid in the weak interacting regime with non-too-low density\cite{?wik J A},
for the polaron dressed by the partially polarized excitations-cloud,
however, the mean-fields approximation sometimes overestimates the interaction effect\cite{Li W}
in the strongly interacting regime (with tightly bound dimers) where the self-trapping, soliton, and breather are harder
to formed than that in the weakly bound pairs (e.g., the BCS superfluid state in the non-Fermi-liquid picture).
The stable repulsive polarons are most likely to be found in the side away from the Feshbach resonance (for gases) 
where polaron energy is large and positive and with $1/k_{F}a \gg 1$ (${\rm ln}(k_{F}a)\ll 0$),
in strong scattering region ($1/k_{F}a\lesssim 1$),
the tightly-bound molecule (within Fermi-liquid picture) could also be found which with negative binding energy 
$E_{b}=-\hbar^{2}/2a^{2}\widetilde{m}<0$ (non-bound state),
as experimentally realized in Ref.\cite{Scazza F,J?rgensen N B},
However, in most case the repulsive polaron is thermodynamically unstable in the 
strong interaction region ($|(k_{F}a_{s})^{-1}|<1$) even at low-temperature.
  While for the attractive polaron,
  whose eigenenergy is negative and the self-energy is real and negative during the whole BEC-BCS crossover,
  can be found in the solid state\cite{Koschorreck M,Mogulkoc A,Modarresi M,Ding Z H,Mogulkoc A2,Falck J P} 
including the topological materials,
  when the condition $k_{F}a_{\psi\phi}\gg 0$ is satisified
($a_{\psi\phi}$ is the scattering length related to the impurity-majority interaction,
which is produced by the attractive potential)
through the, e.g., substrate-related polaronic effect.
Besides, the effects of both the intrinsic and extrinsic (like the electric field-induced Rashba) spin-orbit coupling 
on the polaron in Dirac systems
have also been explored\cite{Mogulkoc A,Modarresi M,Ding Z H,Mogulkoc A2}.
And the spin-orbit coupling is also related to the polaron-molecule transition\cite{Yi W}.
The residue $Z$ (spectral weight), as an important property of the polarons,
which can be measured by the Rabi oscillations\cite{Kohstall C},
is finite even at zero-energy state for the normal Dirac semimetals,
but it vanishes at zero-energy state for the Weyl semimetals which with higher dispersion\cite{Wang J R,Wumulti}
(i.e., the multi-Weyl semimetal).
In the end of the article, we also discuss the pair propagator 
as well as the chiral factor and the self-energy ar finite temperature which exhibit some differences compared to
the zero-temperature case.

\section{Theories}

Different to the non-Fermi-liquid picture in the Dirac/Weyl semimetal,
the impurities of the Fermi gases are mobile as widely studied\cite{Koschorreck M,Scazza F,Fratini E}, while for the 
immobile one the Kondo effect as well as the indirect exchange interactions are considered.
For such mobile impurity,
 the $\delta$-type impurity field (within Born approximation) is valid in measuring the effect of  impurity-interaction.
However, that also cause the vanishing of the spin rotation as well as the intrinsic spin current during the impurity-majority 
scattering (collision) process.
Then, the fixed spin structure guarantees that there exists only the singlet pairing between the impurity and majority particles,
otherwise the triplet pairing exists as discussed in Ref.\cite{Yi W}.
In Fermi gases or the dilute BEC, the interactions between impurity and majority component, and that
between the majority particles needed to taken into account,
such problems are usually dealed by the non-self-consistent many-body $T$-matrix (i.e., the ladder approximation).
Here we consider the interactions between the Fermions (bath) and that between the Bosonic field (the single impurity)
and the Fermionic field.
The difference compared to the normal solid state system\cite{Jian S K,Yang B J,Han S E} is that we are
taking the mobile impurity into account and 
note that we still focus on the single impurity problem since the
Fermi polaron is well defined (with symmetry and easy-to-identify spectral function) in the single-impurity-limit\cite{Koschorreck M}.
At first, we write the microscopic euclidean action as
\begin{equation} 
\begin{aligned}
S=\int d\tau d^{3}r \{\psi^{\dag}[\partial_{\tau}+H_{0}(k)]\psi+\frac{1}{2}\sum_{\alpha=x,y,xz}(\partial_{\alpha}\phi)^{2}
+\frac{1}{2}g_{\psi\psi}\psi^{\dag}\psi\psi^{\dag}\psi+g_{\psi\phi}\psi^{\dag}\psi\phi^{\dag}\phi\},
\end{aligned}
\end{equation}
where $\psi$ and $\phi$ are the Fermionic and Bosonic field, respectively.
$g_{\psi\psi}$ and $g_{\psi\phi}$ are the intraspecies (only around the impurity) and interspecies coupling, respectively,
for the mobile particles,
and
$g_{\psi\phi}$
describes the bare attractive contact interaction strength also.
The contact potential (with Guassian broadening)
which with only the short-range attractive/repulsive interaction 
is used here since the size of pair is much larger than the range of potential,
thus the long-range Coulomb repulsion is reduced here. 
It's important to note that, in solid state,
like the 3D Dirac/Weyl system we discuss here,
the coupling nonzero $g_{\psi\phi}$ requires the nonzero effective mass $m_{\psi}$,
which is the majority component here and we suppose it keeps the up-spin in contrary to the impurity boson throughout this article.
For example, for the bilayer Dirac system, the Fermi energy is similar to that of the Fermi gases due to the existence of the interlayer hopping,
which is $\hbar^{2}k_{F}^{2}/2m$,
while for the Dirac/Weyl semimetal,
the effective mass is absent, and thus the attractive polaron can not formed
except it's dopped.
We here use the coupling constants of the Bose-Fermi mixture system,
\begin{equation} 
\begin{aligned}
g_{\psi\psi}=&\frac{4\pi \hbar^{2}a_{\psi\psi}}{m_{\psi}}<0,\\
g_{\psi\phi}=&[\frac{\widetilde{m}}{2\pi \hbar^{2}a_{\psi\phi}}-\int\frac{d^{3}k}{(2\pi)^{3}}D_{0}(\widetilde{m},k)]^{-1}\\
=&[\frac{\widetilde{m}}{2\pi \hbar^{2}a_{\psi\phi}}-\frac{\widetilde{m}\Lambda}{\pi^{2}}]^{-1}<0,
\end{aligned}
\end{equation}
$\widetilde{m}=m_{\psi}m_{\phi}/(m_{\psi}+m_{\phi})$.
The inversed-energy term $D_{0}(\widetilde{m},k)=\frac{2\widetilde{m}}{k^{2}}$ is proportional to the bare scalar potential (order parameter) propagator in three dimension\cite{wumany}
where the mass term $\widetilde{m}$ is in fact missing in the normal semimetal\cite{Wumulti,Goswami P,Han S E}.
 $\Lambda$ is the momentum cutoff here in order to make the integral convergent.
Note that we set $\Lambda=1$ throughout this paper and it will diverges to infinity when the bare coupling tends to zero.
$a_{\psi\phi}$ is the impurity-majority scattering length which is analogy to the ultraviolet 
divergence.
The $g_{\psi\psi}$ here follows the general definition of the background coupling constant 
which related to the background scattering length $a_{\psi\psi}$.
Indeed, the interaction effect of the polaron system requires the investigation of the effective mass
in contrast to the semimetal,
especially in the strong interaction region ($|(k_{F}a_{s})^{-1}|<1$)
with the obvious renormalization effect and the Fermi-liquid feature.

\section{Results and discussion}
\subsection{$T$-matrix and the self-energy}

The ladder approximation (non-perturbative)
is applicable not only for the imbalanced Fermi gases or nuclear physics, but also for the solid state systems with finite effective mass.
Further, for large-species Fermion system,
the ladder approximation is similar to the leading order $1/N$-expansion.
In thermodynamic limit with $N\rightarrow \infty$,
the self-energy which describes the pairing fluctuation becomes zero\cite{Enss T} 
which implies that the interaction between impurity and majority particles vanishes 
(i.e., without the polaron).
In the strong interacting case,
the self-energy effect as well as the resummation of ladder diagrams (for the forward scattering) are important to be considered.
The non-self consistent $T$-matrix, which does not contains the self-consistency 
of the Green's function, describes the fluctuations in $s$-wave cooper channel.
Firstly, we can write the $T$-matrix between the single impurity and the majority component as
\begin{equation} 
\begin{aligned}
T(p+q,\omega+\Omega)=[\frac{\widetilde{m}}{2\pi \hbar^{2}a_{\psi\phi}}+\Pi(p+q,\omega+\Omega)]^{-1},
\end{aligned}
\end{equation}
where $q$ is the momentum of the majority particle, $p$ is the momentum of the impurity.
Note that for the $T$-matrix here,
we only consider the closed channel scattering (i.e., the bare case),
and without consider the interchanging as well as the spin/valley degrees of freedom.
The term $(p+q)$ is the center-of-mass momentum.
$\Omega\approx \varepsilon_{q\uparrow}=\frac{\hbar^{2}(q^{2}-\mu_{\uparrow}^{2})}{2m_{\psi}}=\frac{\hbar^{2}(q^{2}-k_{F\uparrow}^{2})}{2m_{\psi}}$ is the Fermionic frequency
since we assume the zero-temperature limit,
similarly,
$\omega\approx \varepsilon_{p\downarrow}=\frac{\hbar^{2}(p^{2})}{2m_{\phi}}-\mu_{\downarrow}
=\frac{\hbar^{2}(p^{2})}{2m_{\phi}}-{\rm Re}\ \Sigma(p=0,\omega=0)$ is the Bosonic frequency
where $\Sigma(p,\omega)$ is the impurity self-energy as stated below.
Here $\mu_{\uparrow}\neq \mu_{\downarrow}$ due to the spin-imbalance.
Note that this $T$-matrix is non-self-consistent,
which with the bare impurity propagator and the majority propagator as diagrammatically shown by the Bethe-Salpeter equation
(see, e.g., Ref.\cite{polaron2}).
While the self-consistent $T$-matrix requires the dressed impurity propagator which containing the impurity self-energy effect,
that's more applicable when take into account an infinite number of the majority particles,
in which case its statistics properties emergent
including the imbalance between the two majority species\cite{Pietil? V}.
The Bethe-Salpeter equation about the non-self-consistent many-body $T$-matrix reads
\begin{equation} 
\begin{aligned}
T(p+q,&\square;p+q-k')=V_{0}(p+q,\square;p+q-k')\\           
      &+\sum_{k}V_{0}(p+q,\square;k)G^{\phi}_{0}(p+q-k)G^{\psi}_{0}(\square+k)T(p+q-k,\square+k;p+q-k-k')
\end{aligned}
\end{equation}
where $V_{0}$ are the bare impurity-majority interactions,
specially, $V_{0}(p+q,\square;k)$ is the interaction induced by the polarization operator
(consist of the two bare Green's functions; see Appendix. B).
$k,\ k'$ are the relative momentum.
$G^{\psi}_{0}$ and $G^{\phi}_{0}$ are the bare Fermionic and Bosonic Green's function, respectively,
as presented below.
The symbol $\square$ can be omitted, but we retain it here for the integrity of the above equation.
In the absence of the center-of-mass momentum ($p+q=0$),
the Bethe-Salpeter equation reduced to the Lippmann-Schwinger equation
\begin{equation} 
\begin{aligned}
T(k_{1},k_{2};\omega)=V_{0}(k_{1},k_{2})          
      +\sum_{k_{3}}V_{0}(k_{1},k_{3})\frac{1}{\omega+i0-2\varepsilon_{k_{3}}}T(k_{3},k_{2},\omega).
\end{aligned}
\end{equation}

The impurity-majority pair propagator reads
\begin{equation} 
\begin{aligned}
\Pi(p+q,\omega+\Omega)
=\int\frac{d^{3}k}{(2\pi)^{3}}\int\frac{d\nu}{2\pi}
                             G_{0}^{\psi}(\nu,k)G_{0}^{\phi}(\omega+\Omega-\nu,p+q-k)-\frac{\widetilde{m}\Lambda}{\pi^{2}},
\end{aligned}
\end{equation}
where $G^{\psi}_{0}(\nu,k)=[\nu+i0^{+}-\frac{\hbar^{2}k^{2}}{2m_{\psi}}+\mu_{\uparrow}]^{-1}$ is the noninteracting 
(in the absence of a condensate and the long-range Coulomb interaction) majority particle (Fermion) propagator 
and $G^{\phi}_{0}(\omega+\Omega-\nu,p+q-k)=[\omega+\Omega-i\nu+2i0^{+}-\frac{\hbar^{2}(p+q-k)^{2}}{2m_{\phi}}+\mu_{\downarrow}]^{-1}$ 
is the bare impurity propagator (not the scalar-field one).
In this majority particle propagator we ignore the perturbation from the single impurity to the Fermi sea.

For a comparative study,
we at first discuss the self-energy in the general diluted Fermi gases,
which is usually referred to as the polaronic binding energy\cite{Li W} or the molecule binding energy\cite{Yu Z Q,Koschorreck M}.
This self-energy can be obtained by the following impurity-majority interaction Hamiltonian
\begin{equation} 
\begin{aligned}
H_{{\rm int}}=\begin{pmatrix}
g_{\psi\phi}|\Psi_{\psi}|^{2} & 0\\
0& g_{\psi\phi}|\Psi_{\phi}|^{2}
\end{pmatrix},
\end{aligned}
\end{equation}
and the related the interaction energy 
\begin{equation} 
\begin{aligned}
\varepsilon_{\psi\phi}=n_{F}\int d^{3}R \Psi(R)[\frac{-\hbar^{2}\nabla_{R}^{2}}{2\widetilde{m}}+U_{\psi\phi}]\Psi^{\dag}(R),
\end{aligned}
\end{equation}
where $\Psi(R)=(\Psi_{\psi}(R),\Psi_{\phi}(R))$ is the normalized wave function.
$n_{F}=\int\frac{d^{3}q}{(2\pi)^{3}}\int\frac{d\Omega}{2\pi}G_{F}(q,\Omega)e^{i0^{+}\Omega}$ is the numerical density of the Fermions,
where $G_{F}(q,\Omega)$ 
is the dressed (full) Green's function unlike the above ones,
and gives the actural Fermion dispersion.
There exists the constrains
\begin{equation} 
\begin{aligned}
4\pi n_{F}\int^{\Lambda}_{0}dR R^{2}\Psi(R)\Psi^{\dag}(R)=&{\bf I},\\
|\Psi(R\ge \Lambda)|=&1,
\end{aligned}
\end{equation}
where ${\bf I}$ is the unitary matrix.
Base on the many-body scattering theory at low-temperature,
where we consider only the $s$-wave scattering,
the $T$-matrix in Fermi (or Bose) gases is usually self-consistent,
i.e.,
it's a two-channel $T$-matrix\cite{Massignan P,Bruun G M}
while in our model with Dirac Fermions,
the $T$-matrix contains only the closed channel (i.e., the bare one)
which is thus non-self-consistent,
since we consider only a finite number of the majority particles.

To study the many-body effect, the non-self-consistent $T$-matrix is similar but not exactly like the leading order $1/N$ expansion,
since it ignore the dynamical screening effect.
For this reason, the non-self-consistent $T$-matrix is more like the leading-order loop expansion with GV approximation rather than the leading order $1/N$ expansion with GW approximation.
The related studies are also reported in Refs.\cite{Enss T,Rath S P,Nikolić P}.
Besides, the validity of the non-self-consistent $T$-matrix in studing the BCS-BEC crossover has also been verified
\cite{Tsuchiya S,Haussmann R}.

In our model,
the self-energy about the interaction between mobile impurity and the bath reads
\begin{equation} 
\begin{aligned}
\Sigma(p,\omega)=\sum_{q<k_{F}}\frac{N_{F}(\Omega)}
{\frac{\widetilde{m}}{2\pi \hbar^{2}a_{\psi\phi}}+\Pi(p+q,\omega+\Omega)},
\end{aligned}
\end{equation}
and the numerator can be replaced by $\theta(k_{F}-q)$ at zero-temperature limit, where $\theta$ stands the step function.
Note that this self-energy expression describes only the region around the single impurity (the attractive polaron)
which is small but not localized (since the impurity is mobile).
It's different to the Fermi gases that,
the self-energy of the impurity does not contains the condensate density as well as the condensate-related spin fluctuation
and the pair propagator contains the chiral factor $F_{ss'}\ (s,s'=\pm 1)$ (the wave function overlap) which 
suppresses the backscattering and is absent in the 2D electron gas.
The chiral factor here is indeed observble in the polarons formed in the Dirac system\cite{Kandemir B S}.
While for 2D electron gas,
$F^{ss'}=F^{+}=1$ and contains only the intraband contribution,
except at a quantum Hall setup with strong magnetic filed as report in Ref.\cite{Bocquillon E}.
We can also see that,
in the narrow gap limit,
the pair propagator reduced to the well known dynamical polarization,
and $T=\Pi^{-1}$.
In the surface of Dirac system,
since away from the condensed phase, the condensate density vanishes but the related pairing fluctuations remain as long as 
$g_{\psi\phi}\neq 0$, i.e., the pairing instability exists (especially when the spin-orbit coupling turns on\cite{Yi W})
even in the case of $g_{\psi\psi}=0$.

\subsection{Pair propagator at zero-temperature limit}
At zero-temperature limit,
the above pair propagator can be written as
\begin{equation} 
\begin{aligned}
\Pi(p+q,\omega+\Omega)
=&  \int\frac{d^{3}k}{(2\pi)^{3}}
    \frac{1-N_{F}(\varepsilon_{k\uparrow}-\mu_{\uparrow})}
{-\omega-i0-\Omega+\varepsilon_{k\uparrow}+\varepsilon_{p+q-k\downarrow}}F_{ss'}-\int\frac{d^{3}k}{(2\pi)^{3}}
D_{0}(\widetilde{m},k)\\
=&  \int\frac{d^{3}k}{(2\pi)^{3}}
    \frac{N_{F}(\varepsilon_{k\uparrow}-\mu_{\uparrow})-1}
{\omega+i0+\Omega-\varepsilon_{k\uparrow}-\varepsilon_{p+q-k\downarrow}}F_{ss'}-\int\frac{d^{3}k}{(2\pi)^{3}}
D_{0}(\widetilde{m},k)\\
=&\frac{4\pi}{(2\pi)^{3}}\int^{\Lambda}_{k_{F}}
    \frac{-k^{2}\theta(k-k_{F})}{\omega+i0+\varepsilon_{q\uparrow}-\varepsilon_{k\uparrow}-\varepsilon_{p+q-k\downarrow}}F_{ss'}dk
-\frac{\widetilde{m}\Lambda}{\pi^{2}},
\end{aligned}
\end{equation}
where $N_{F}$ is the Fermi-distribution function and it appears only in the presence of nonzero center-of-mass momentum.
Base on this expression as well as the above self-energy, we can obtain that,
the polaron self-energy increase with the increasing Dirac-mass term or the coupling parameter $g_{\psi\phi}$,
however, there is an exception:
when the intrinsic spin-orbit coupling (not the extrinsic one) is present,
then the increase of $g_{\psi\phi}$ will reduce the self-energy since it greatly reduce the Dirac-mass\cite{Mogulkoc A}.
As we can see, although the chiral factor $F_{ss'}$ is contained, 
it's in fact dominated by the backscattering in bilayer Dirac system\cite{Adam S} as in the 2D electron gas,
so next we at first discuss the case in the absence of chiral factor ($F_{ss'}=1$).
For $F_{ss'}=1$, we obtain
\begin{equation} 
\begin{aligned}
\Pi(p+q,\omega+\Omega)
=&\frac{4\pi}{(2\pi)^{3}}\frac{2}{\hbar^{3}(m_{\psi}+m_{\phi})^{2}}
\left[-\hbar km_{\psi}m_{\phi}(m_{\psi}+m_{\phi})\right.\\
&\left.+
\frac{m_{\psi}\sqrt{m_{\phi}}A\ {\rm arctan}[\frac{\hbar (-m_{\psi}(p+q)+k(m_{\phi}+m_{\psi}))}
{\sqrt{m_{\phi}}\sqrt{\hbar^{2}m_{\psi}p(p+2q)+2m_{\psi}^{2}(-\omega-i0+\mu_{\uparrow})
 -\hbar^{2}q^{2}m_{\phi}+2m_{\phi}m_{\phi}(-\omega-i0+\mu_{\uparrow}) }}]}{B}+C
\right]\bigg|_{k_{F}}^{\Lambda}\\
&-\frac{\widetilde{m}\Lambda}{\pi^{2}},
\end{aligned}
\end{equation}
where we define
\begin{equation} 
\begin{aligned}
A=&-2m_{\psi}(-\omega-i0+\mu_{\uparrow})m_{\phi}(m_{\psi}+m_{\phi})+\hbar^{2}(m^{2}_{\psi}(p+q)^{2}-m_{\psi}p(p+2q)m_{\phi}+q^{2}m_{\phi}^{2}),\\
B=&\sqrt{\hbar^{2}m_{\psi}p(p+2q)+2m_{\psi}^{2}(-\omega-i0+\mu_{\uparrow})-\hbar^{2}q^{2}m_{\phi}+2m_{\psi}m_{\phi}(-\omega-i0+\mu_{\uparrow})},\\
C=&-\hbar m_{\psi}^{2}(p+q)m_{\phi}{\rm ln}[2m_{\psi}m_{\phi}(-\omega-i0+\mu_{\uparrow})\\
&+\hbar^{2}(-2km_{\psi}(p+q)+m_{\psi}(p+q)^{2}-q^{2}m_{\phi}+k^{2}(m_{\psi}+m_{\phi}))].
\end{aligned}
\end{equation}
We found that, the logarithmic term (within the expression of $C$) vanish in the limit of $p\rightarrow 0,q\rightarrow 0$,
and in such case, if we set $m_{\phi}=0.1$,
$m_{\psi}=0.01$, then the pair propagator can be simplified as (with positive impurity frequency $\omega$)
\begin{equation} 
\begin{aligned}
\Pi(p+q,\omega+\Omega)\approx -0.018\Lambda+0.0025\sqrt{-i\omega}\ {\rm arctan}\frac{7.42\Lambda}{\sqrt{-i\omega}}.
\end{aligned}
\end{equation}
Here the renormalized mass $\widetilde{m}=0.009$ is very small and has $m_{\phi}\gg m_{\psi}$
which is appropriate for the scenario of the Dirac system with a dropped impurity.
That also results in a small value of the bare scalar potential propagator $D_{0}(\widetilde{m},k)$.
We want to note that,
during the calculation of above pair propagator,
the Sokhotski-Plemelj theorem is applicable for the analytic continuation of the imaginary frequency which 
with very small imaginary part.
For $\mu_{\uparrow}\neq 0$,
the above expression can be rewritten as
\begin{equation} 
\begin{aligned}
\Pi(p+q,\omega+\Omega)\approx -0.018\Lambda+0.0025\sqrt{-i\omega+\mu_{\uparrow}}\ {\rm arctan}\frac{7.42\Lambda}{\sqrt{-i\omega+\mu_{\uparrow}}},
\end{aligned}
\end{equation}
In Fig.1, we show the pair propagator at non-chiral case
as a function of the impurity momentum $p$ and majority particle momentum $q$.
The chemical poential of the majority particle is setted as zero here, i.e., the noninteracting Fermionic bath.
In Fig.2, the pair propagator with chemical potential $\mu_{\uparrow}=0.5$ is presented.
The intraspecies coupling $g_{\phi\phi}$ is related to both the effective mass (the electron interaction) 
and the chemical potential $\mu_{\uparrow}$,
and it's important in keeping the stability.
Then by comparing the Fig.1 and Fig.2,
we can easily found that the system in Fig.2 is more stable with finite chemical potential $\mu_{\uparrow}$
which exhibits less peaks even though we have employed a more dense mesh compared to the Fig.1,
that's agree with our previous inference.
While in Fig.3, we present the pair propagator at $p=q=0$.
We found that at $p=q=0$, the real part of the pair propagator is negative,
while the imaginary part contains both the positive and negative values.
This figure
is very similar to our previous result\cite{ele}
about the disorder-induced self-energy in a gapless parabolic Dirac system,
which still has the negative real part but complete positive imaginary part.
This similarity can be explained by the expression of impurity self-energy within the polaron
which describing the interactions between the impurity and the majority particles:
\begin{equation} 
\begin{aligned}
\Sigma(p,\omega)=\int^{k_{F}}_{0}\frac{d^{3}q}{(2\pi)^{3}}\int\frac{d\Omega}{2\pi}
N_{F}(\varepsilon_{q\uparrow})T(p+q,\omega+\Omega)G^{0}_{\psi}(q,\Omega).
\end{aligned}
\end{equation}
A full description of the self-energy can also be obtained by replacing the Green's function within above expression
by the full one (through the Dyson equation):
$G_{\psi}(q,\Omega)=[\Omega+i0-\varepsilon_{q\uparrow}+\mu_{\uparrow}-\Sigma(q,\Omega)]^{-1}$
where $\Sigma(q,\Omega)$ is the self-energy of the majority Fermions,
which can be similarly written as
\begin{equation} 
\begin{aligned}
\Sigma(q,\Omega)=-\int\frac{d^{3}p}{(2\pi)^{3}}\int\frac{d\omega}{2\pi}
N_{F}(\varepsilon_{q\uparrow})T(p+q,\omega+\Omega)G^{0}_{\phi}(p,\omega).
\end{aligned}
\end{equation}

For the case of $F_{ss'}\neq 1$, i.e., with not only the backscattering,
we obtain for intraband transition
\begin{equation} 
\begin{aligned}
\Pi(p+q,\omega+\Omega)
=&-\frac{4\pi}{(2\pi)^{3}}
  \frac{m_{\psi}m_{\phi}}{4\hbar^{3}(m_{\psi}+m_{\phi})^{3}}
\left[8\hbar k(m_{\psi}+m_{\phi})\right.\\
&\left.+
       \frac{  A\ {\rm arctan}
[\frac{\hbar (-m_{\psi}(p+q)+k(m_{\phi}+m_{\psi}))}
{\sqrt{m_{\phi}}\sqrt{\hbar^{2}m_{\psi}p(p+2q)+2m_{\psi}^{2}(-\omega-i0+\mu_{\uparrow})
 -\hbar^{2}q^{2}m_{\phi}+2m_{\phi}m_{\phi}(-\omega-i0+\mu_{\uparrow}) }}]}{B}+C
\right]\bigg|_{k_{F}}^{\Lambda}\\
&-\frac{\widetilde{m}\Lambda}{\pi^{2}},
\end{aligned}
\end{equation}
where we define
\begin{equation} 
\begin{aligned}
A=&-16m_{\psi}(-\omega-i0+\mu_{\uparrow})m_{\phi}(m_{\psi}+m_{\phi})\\
    &+\hbar^{2}(m^{2}_{\psi}(8p^{2}+16pq+7q^{2})^{2}-2m_{\psi}(4p^{2}+8pq+q^{2})m_{\phi}+7q^{2}m_{\phi}^{2})\\
    &+\hbar^{2}q^{2}(m_{\psi}+m_{\phi})^{2}{\rm cos}(2a),\\
B=&\sqrt{m_{\phi}}\sqrt{\hbar^{2}m_{\psi}p(p+2q)+2m_{\psi}^{2}(-\omega-i0+\mu_{\uparrow})-\hbar^{2}q^{2}m_{\phi}+2m_{\psi}m_{\phi}(-\omega-i0+\mu_{\uparrow})},\\
C=&8\hbar m_{\psi}(p+q)
    {\rm ln}[2m_{\psi}m_{\phi}(-\omega-i0+\mu_{\uparrow})\\
&+\hbar^{2}(-2km_{\psi}(p+q)+m_{\psi}(p+q)^{2}-q^{2}m_{\phi}+k^{2}(m_{\psi}+m_{\phi}))],
\end{aligned}
\end{equation}
where $b$ is the angle between $k$ and $k'$ and $a$ is the angle between $k$ and $q$:
\begin{equation} 
\begin{aligned}
{\rm cos}\ b&=\frac{k+q{\rm cos}a}{\sqrt{k^{2}+q^{2}+2kq{\rm cos}a}}=1-\frac{q^{2}{\rm sin}^{2}a}{2k^{2}}+O(q^{3}),
\end{aligned}
\end{equation}
i.e., we suppose the chiral factor is independent of the impurity momentum $p$.
Similarly, for interband transition,
we have
\begin{equation} 
\begin{aligned}
\Pi(p+q,\omega+\Omega)
=&-\frac{4\pi}{(2\pi)^{3}}\\
&\times       \frac{ m_{\psi}q^{2}\sqrt{m_{\phi}}{\rm sin}^{2}a\ {\rm arctan}
[\frac{\hbar (-m_{\psi}(p+q)+k(m_{\phi}+m_{\psi}))}
{\sqrt{m_{\phi}}\sqrt{\hbar^{2}m_{\psi}p(p+2q)+2m_{\psi}^{2}(-\omega-i0+\mu_{\uparrow})
 -\hbar^{2}q^{2}m_{\phi}+2m_{\phi}m_{\phi}(-\omega-i0+\mu_{\uparrow}) }}]}
{2\hbar
\sqrt{\hbar^{2}m_{\psi}p(p+2q)+2m_{\psi}^{2}(-\omega-i0+\mu_{\uparrow})-\hbar^{2}q^{2}m_{\phi}+2m_{\psi}m_{\phi}(-\omega-i0+\mu_{\uparrow})}
}\bigg|_{k_{F}}^{\Lambda}\\
&-\frac{\widetilde{m}\Lambda}{\pi^{2}},
\end{aligned}
\end{equation}
The results are presented in Figs.4-5,
as we can see,
the fluctuation of the pair propagator increase when angle $a=\pi/4$ turns to $a=\pi/2$,
and decrease when $a=\pi/2$ turns to $a=3\pi/4$,
which implies that the fluctuation of $\Pi(p+q,\omega+\Omega)$ is proportional to the $({\rm cos}\ a)$.

We note that, for the case of large-$\mu_{\uparrow}$ in Fermi-liquid picture,
the Fermion interaction is dominated by the short-range one,
i.e., the Hubbard interaction,
due to the strong screening effect.
Then the strong spin fluctuation as well as the particle-hole fluctuation
are possible to build the bipolaron as predicted recently\cite{Camacho-Guardian A}.
The quasiparticle properties can be view clearly by the spectral function
which measures the propability of exciting or removing a (quasi)particle at a certain momentum.
It's directly that the spectral function as a function of impurity momentum will exhibits a similar dispersion with that of the
eigenenergy of the impurity (i.e., the quadratic dispersion here),
and it increases faster than the molecule one\cite{Rath S P},
which with much smaller value of $k_{F}a_{\psi\phi}$.
We plot in Fig.6 the spectral function (of the hole states) at zero impurity momentum ($p=0$) of the Dirac system 
with and without the Hubbard interaction.
The many-electron effect is contained in this figure as can be seen from the broaden peaks.
The peaks will be pronounced (not the $\delta$-type one) in the self-consistent Green's function beyond the mean-field theory\cite{Hassaneen K S A}.
As we can see, the spectral function decreases 
and the fluctuation (of probability) is enhanced when the short-range Hubbard interaction increases
which implies that the stability of system is enhanced (due to the decreasing electron number),
and the sensitivity of the
spectral function to the impurity frequency is also increased.
Next we consider the particle spectral function containing the many-body effect,
\begin{equation} 
\begin{aligned}
A(p,\Omega)=\frac{|{\rm Im}\Sigma(p,\omega)|}
{(\Omega-{\rm Re}\Sigma(p,\omega)-\frac{\hbar^{2}p^{2}}{2m_{\phi}}+\mu_{\downarrow})^{2}+({\rm Im}\Sigma(p,\omega))^{2}}.
\end{aligned}
\end{equation}

Within one-particle-hole approximation,
which is valid
according to the Monte Carlo calculation and the 
experimental results due to the destructive interference in the presence of more than one particle-hole part\cite{Combescot R2},
the variational wave function reads
\begin{equation} 
\begin{aligned}
|\psi\rangle=
\psi_{0}c_{p\downarrow}^{\dag}|0\rangle_{\uparrow}+\sum_{k>k_{F},q<k_{F}}\psi_{kq}c_{p+q-k,\downarrow}^{\dag}
c_{k,\uparrow}^{\dag}c_{q,\uparrow}|0\rangle_{\uparrow},
\end{aligned}
\end{equation}
where $|0\rangle_{\uparrow}=\Pi_{k<k_{F}}c_{k\uparrow}^{\dag}|{\rm vac}\rangle$ is the ground state of the majority particles\cite{Combescot R}
and $|{\rm vac}\rangle$ is the vacuum electron state.
The first term in the right-hand-side of above equation describes the free impurity which assumed is that totally delocalized.
$k$ is the momentum of a majority-particle scattered out of fermi surface, and
$q$ is the momentum of a majority-particle before scattering.
Through the normalization condition $\langle\psi|\psi\rangle=1$, we have,
after minimizing the total energy,
\begin{equation} 
\begin{aligned}
\psi_{kq}=&\psi_{0}\frac{T(p+q,\omega+\Omega)}{\omega-\varepsilon_{p+q-k,\downarrow}-\varepsilon_{k,\uparrow}+\varepsilon_{q,\uparrow}},\\
\psi_{0}=&\frac{1}{\sqrt{1+\sum_{k>k_{F},q<k_{F}}(\frac{\psi_{kq}}{\psi_{0}})^{2}}}.
\end{aligned}
\end{equation}
At zero energy limit, the quasiparticle residue of the usual Dirac Fermions remains finite 
(here we assuming a noninteracting initial state)
and thus the coefficients $\psi_{0}$
and $\psi_{kq}$ remain finite too,
while for the multi-Weyl semimetal, the residue vanishes at zero-energy and then $\psi_{0}=\psi_{kq}=0$.
Foe zero momentum ($q=0$) with the lowest dispersion,
the impurity self-energy becomes,
\begin{equation} 
\begin{aligned}
\Sigma(p,\omega)=\frac{1}{S}\sum_{\mu_{\uparrow}}\frac{1}{\frac{\widetilde{m}}{2\pi\hbar^{2}a_{\psi\phi}}+\Pi(\omega+\Omega,p)}
\frac{1}{\omega+i0+\mu_{\uparrow}},
\end{aligned}
\end{equation}
with $\Pi(\omega+\Omega,p)=\Pi(\omega+\Omega,p,q=0)$, and 
$S$ is the volume of the space where all the chemical potential around the polaron are taken into account.
Now the center of mass is just $p$.
That's also agree with the results of the Fr\"{o}hlich polaron model\cite{Mogulkoc A2,Grusdt F} in long-wavelength limit which
with the strong electron-phonon coupling as well as the observable optical excitations\cite{Falck J P}.

For simplicity, we further set $\mu_{\uparrow}=0$ and the Fermi frequency $\Omega=1$ (which is possible in the zero-temperature limit),
then the polaron self-energy becomes
$\Sigma(p,\omega)=T(p,\omega)G^{0}_{\psi}(0,0)$ where $G^{0}_{\psi}(0,0)=1$,
thus
the above self-energy can be written as
\begin{equation} 
\begin{aligned}
\Sigma(p,\omega)^{-1}=&
-\frac{2m_{\psi}m_{\phi}\left(\hbar k(m_{\psi}+m_{\phi})+\frac{A\ {\rm atanh}(\frac{\hbar m_{\psi}p-\hbar k(m_{\psi}+m_{\phi})}{\sqrt{m_{\psi}m_{\phi}\sqrt{-\hbar^{2}p^{2}+2i0(m_{\psi}+m_{\phi})+2\omega(m_{\psi}+m_{\phi})}}})}{B}
+C\right)}
{\hbar^{3}(m_{\psi}+m_{\phi})^{2}}\Bigg|_{k}\\
&+\frac{\widetilde{m}}{2\pi\hbar^{2}a_{\psi\phi}},
\end{aligned}
\end{equation}
where
\begin{equation} 
\begin{aligned}
A=&\sqrt{m_{\psi}}(\hbar^{2}p^{2}(m_{\psi}-m_{\phi})+2(i0+\omega)m_{\phi}(m_{\psi}+m_{\phi})),\\
B=&\sqrt{m_{\phi}}\sqrt{-\hbar^{2}p^{2}+2i0(m_{\psi}+m_{\phi})+2\omega(m_{\psi}+m_{\phi})},\\
C=&\hbar m_{\psi}p{\rm ln}(2m_{\psi}(\omega+i0)m_{\phi}-\hbar^{2}(-2kpm_{\psi}+m_{\psi}p^{2}+k^{2}(m_{\psi}+m_{\phi}))).
\end{aligned}
\end{equation}
The self-energy is shown in Fig.7(a).
Then by substituting the above self-energy to the Eq.(22),
we obtain the corresponding particle spectral function as shown in the Fig.7(b),
where we set $\omega>0\ (\omega>\mu_{\uparrow})$ to make sure the spectral function here describes only the particle states.
Through Fig.7(b),
the total density of states can be obtained by integrating over the $p$-axis,
while the occupation probability\cite{Hassaneen K S A} can be obtained by integrating over the $\omega$-axis.

\section{Summary}

In summary, we investigate the properties of the attractive polaron 
formed by a single impurity immersed into the the Fermi bath in a Dirac system
at zero-temperature limit.
Within the pair propagator,
a bare scalar potential (order parameter) term is contained which is benefit to the convergence of integral at large momentum,
we still use the ultraviolet cutoff 
which is setted as $\Lambda=1$ throughout the paper.
Besides, althought the pair propagator is the bare one,
it's renormalized by the bare scalar potential term as shown in the main text.
The 
hole spectral function under zero and nonzero short-range Hubbard interaction is shown.
And the particle spectral function of the polaron in the presence of quantum many-body effect
is also discussed in this paper,
which exhibits obvious fluctuation.
  That's in contrast to the spectral function of the immobile impurity in intrinsic graphene which with linear dispersion
but not the parabolic one as we discussed.
  The later case is studied in Ref.\cite{Skrypnyk Y V}
where we see that the fluctuations of the spectral function can't be found,
and the peak broadening is also much smaller.
However, that doesn't means that the intrinsic graphene with linear dispersion can not forms the polaron,
as founded within the electron-phonon coupling systems\cite{Kandemir B S}.

In this paper, we mainly discuss the polaron in the surface of the three-dimensional Dirac system,
the method reported here can also be applied to the two-dimensional Dirac systems
with finite effective mass.
In the numerical simulations,
we studied the instability of the pair propgator and the spectral function
by establishing a very small-renormalized mass and by varying the chemical potential.
We found that,
the non-chiral pair propagator becomes more stable when with finite chemical potential,
and for chiral pair propagator, it's proportional to the ${\rm cos}\ a$
where $a$ is the angle between the wave vector before scattering and the scattering wave vector.
The many-body effect is also analyzed through the study of spectral function.
We note that, in the calculation of the renormalized interspecies coupling which is cutoff-independent,
the bare scalar potential propagator-like factor $D_{0}\sim 1/k^{2}$ is used\cite{Li W,Rath S P,Fratini E},
which is in fact a inversed kinetic energy term, and thus it is independent of the dimension.
Although the $T$-matrix approximation here takes into account the pairing interaction 
with the leading instability even in the presence of weak intraspecies interactions (the $p$-wave interaction), 
it is indeed a nonperturbation theory which is evidened by the absence of the self-consistency
(i.e., the Coulomb induced exchange self-energy is the Hartree-Fock type and in lack of the dynamical dielectric function),
thus the energy is unconserving 
and the quasiparticle weigth is lower than the random phase approximation (RPA) theory 
(with the dynamical screening) or the partial self-consistent one (with static screened interaction or dielectric function)\cite{Holm B}.
Recently, the formation and properties of the attractive polaron formed in a two-dimensional semi-Dirac system is reported by us\cite{polaron2}, 
where the anisotropic effective masses distribution takes an important role,
and we approximate the dispersion as the parabolic anisotropic one 
similar to the dispersion of the plasmon-polaron formed by the phosphorene locate on polar substrates\cite{Saberi-Pouya S}.
Besides, the $p$-wave scattering of the polaron system is also studied in topological superfluid and the weak-coupled BEC recently\cite{Kinnunen J J,Qin F}.
The attractive polaron as a quasiparticle can be observed experimently through the momentum-resolved photoemission spectroscopy\cite{Koschorreck M}.
For Bose-Hubbard model in superfluid phase, 
the Bose polaron with a spin impurity can be created by the off-resonance laser and microscope objective\cite{Fukuhara T,Weitenberg C}.
Specially, at half-filling ($\mu=0$) where the electron density equals 1 and the on-site Hubbard U is much larger than the mobility of impurity, the 
spin impurity is localized and in this case the non-self-consistent $T$-matrix approximation has high accuracy 
due to the weak dynamical screening effect from the carriers.
In one-dimensional geometry, this experiment also provides a platform to explore the other polaronic physics like the propagation velocity affected by the self-trapping effect,
which implies that the polaronic effect can emergents also in the superconductors or the Mott insulators\cite{Endres M}.
The self-trapping effect will becomes more obvious at finite temperature due to the emergent electron-phonon coupling.
While at low-temperature limit (e.g., $<1\mu K$)
the magnetic or electric trapping can be applied to the molecule cloud or the hyperfine states 
(can be treated as the species as we discuss in above) of the alkali atoms,
to design the quantum memory setups in the quantum circuit\cite{Rabl P}.
For solid state like the Dirac system as we discussed, in the presence of, 
e.g., the  separable $s$-wave potential\cite{Gaul C}, 
the $s$-wave scattering as well as the elastic scattering can be treated as dominating at low-temperature limit, 
and the two-body Lippmann-Schwinger equation is still valid in obtaining the coupling parameters and the $T$-matrix, 
and in fact for impurity and the particle-hole part (excited by the quantum fluctuation) with energies similar to the same Dirac cone, 
the scattering can be treated as the intravalley one, which can help us to deal with the multichannel problem.

\section{Appendix}

\subsection{A: Variational approach in mean-field approximation for isotropic lattice}

For Bose polaron in BEC,
the method of mean-field approximation is valid in the presence of the weak on-site Boson-Boson or Boson-Fermion (impurity)
interaction, i.e., the dilute BEC,
and certainly,
the physical parameters like the lattice parameter or the interaction strength
can be controlled by the Feshbach technique,
and the strong-interaction regime can also reached by this method.
Here we use the variational approach base on Gaussian variational ansatz and the Lagrangian optimization.
The variational approach can be generalized by the differential of the matrix element\cite{Li W}
\begin{equation} 
\begin{aligned}
\frac{\partial \langle \Psi|H-E|\Psi\rangle}{\partial (i\psi^{*})}=0,
\end{aligned}
\end{equation}
where $\Psi$ is the trial wave function including the interaction effect,
$H$ is the effective Hamiltonian of the discussing system,
$E$ is the Lagrange multiplier which gives the local minimal energy,
$\psi$ is the real components.
We make the mean-field approximation to the Grassmann field which written as $c_{j}$ at site $j$,
then the Lagrangian reads
\begin{equation} 
\begin{aligned}
L=\sum_{j}i\frac{\partial H^{MF}}{\partial (ic_{j}^{*})}c_{j}^{*}-H^{MF},
\end{aligned}
\end{equation}
where $H^{MF}$ is the mean-field Hamiltonian,
the Grassmann field is treated as a dynamical Gaussian profile as
\begin{equation} 
\begin{aligned}
c_{j}=\frac{\sqrt{2}}{r\sqrt{\pi}}{\rm exp}[-\frac{(j-c)^{2}}{r^{2}}+ik(j-c)],
\end{aligned}
\end{equation}
where $c$ and $k$ is the coordinate and momentum of the center of wave package, respectively,
$r$ is the width of wave package.

Then base on the Euler-Lagrangian relation 
\begin{equation} 
\begin{aligned}
\frac{\partial L}{\partial c}-\frac{d}{dt}\frac{\partial L}{\partial \dot{c}}=-m\ddot{c}=0,
\end{aligned}
\end{equation}
we can obtain 
\begin{equation} 
\begin{aligned}
c={\rm sin}k\ e^{-\frac{1}{2r^{2}}}t,\\
m=\frac{1}{\hbar^{2}}{\rm cos}k\ e^{-\frac{1}{2r^{2}}},
\end{aligned}
\end{equation}
In the absence of the external potential,
the effective Hamiltonian is independent of $c$,
\begin{equation} 
\begin{aligned}
\frac{\partial H^{MF}}{\partial c}=&0,\\
\frac{\partial H^{MF}}{\partial \dot{c}}=&\left[-\frac{A}{r^{2}}-{\rm cos}k\ e^{\frac{-1}{2r^{2}}}\frac{1}{r^{3}}\right]
\frac{r^{3}}{{\rm sin}k\ e^{\frac{-1}{2r^{2}}}},
\end{aligned}
\end{equation}
where the effective coupling parameter $A=U/(4J\sqrt{\pi})$ as a ratio between the on-site interaction $U$ and the tunneling strength $J$,
the mean-field Hamiltonian here reads $H^{MF}=\frac{A}{r}-{\rm cos}e^{\frac{-1}{2r^{2}}}$.
Here we note that the critical value of effective coupling parameter $A$ for the 
self-trapping, soliton, and breather are not continued during the BEC-BCS crossover,
 unlike the attractive self-energy beyong the Hartree-Fock approximation\cite{Li W}.
In the strong interacting case,
the electron may become self-trapped and with localized wave package
characterized by a diverging effective mass $m$.

\subsection{B: Derivation of Pair propagator at finite temperature}

Firstly, for the case of finite temperature,
the fermionic and bosonic Matsubara frequencies read $\Omega=(2n+1)\pi T$, $\nu=(2n'+1)\pi T$ and $\omega=2m\pi T$ 
($n,n',m$ are integer numbers), respectively,
which are discrete variables.
The pair propagator can be written as
\begin{equation} 
\begin{aligned}
\Pi(p+q,\omega+\Omega)
=\int\frac{d^{3}k}{(2\pi)^{3}}[\sum_{n'}\frac{T}{V}
                             G_{0}^{\psi}(\nu,k)G_{0}^{\phi}(\omega+\Omega-\nu,p+q-k)-\frac{2\widetilde{m}}{k^{2}}],
\end{aligned}
\end{equation}
where we consider the single band model and regard the Green's functions as the only eigenvalue of the matrix,
thus we need't write the term (consists of the two Green's functions) as 
$S={\rm Tr}[\sigma_{0}G_{0}^{\psi}(\nu,k)\sigma_{0}G_{0}^{\phi}(\omega+\Omega-\nu,p+q-k)]$
as emerges in the one-loop polarization.
Here we define (hereafter we set $\hbar=1$)
\begin{equation} 
\begin{aligned}
G_{0}^{\psi}(\nu,k)=&\frac{1}{i\nu-\frac{k^{2}}{2m_{\psi}}+\mu_{\uparrow}},\\
G_{0}^{\phi}(\omega+\Omega-\nu,p+q-k)=&\frac{1}{i\omega+i\Omega-i\nu-\frac{(p+q-k)^{2}}{2m_{\phi}}+\mu_{\downarrow}}.
\end{aligned}
\end{equation}
The summation over Matsubara frequencies ($i\nu$) can be calculated as
\begin{equation} 
\begin{aligned}
\sum_{n'=-\infty}^{\infty}\frac{1}{i(2n'+1)\pi T-a}\frac{1}{-i(2n'+1)\pi T-b}=
\frac{{\rm tanh}\frac{b}{2T}+{\rm tanh}\frac{a}{2T}}
{2T(a+b)},
\end{aligned}
\end{equation}
where $a\equiv \varepsilon_{k\uparrow}=\frac{k^{2}}{2m_{\psi}}-\mu_{\uparrow}$,
  $b\equiv -i\omega-i\Omega+\frac{(p+q-k)^{2}}{2m_{\phi}}-\mu_{\uparrow}$,
and
that can also be rewritten as
\begin{equation} 
\begin{aligned}
\sum_{n'=-\infty}^{\infty}\frac{1}{i(2n'+1)\pi T-a}\frac{1}{-i(2n'+1)\pi T-b}=
\frac{N_{B}(a+b)}{N_{F}(a)N_{F}(b)T(a+b)},
\end{aligned}
\end{equation}
where $N_{B}(x)=1/(e^{x/T}-1)$ is the Bose distribution function
and $N_{F}(x)=1/(e^{x/T}+1)$ is the Fermi distribution function.
For small $a$ and $b$, i.e., in the limit of small energy and small frequency,
we approximate ${\rm tanh}(x)\approx x-\frac{x^{3}}{3}+O(x^{5})$,
then the pair propagator becomes
\begin{equation} 
\begin{aligned}
\Pi(p+q,\omega+\Omega)
=&4\pi\frac{1}{(2\pi)^{3}}\int^{\Lambda}_{k_{F}} k^{2}dk
\left[\frac{{\rm tanh}\frac{b}{2T}+{\rm tanh}\frac{a}{2T}}
{2T(a+b)}-\frac{2\widetilde{m}}{k^{2}}\right]\\
=&4\pi\frac{1}{(2\pi)^{3}}\int^{\Lambda}_{k_{F}} k^{2}dk
\left[\frac{\frac{b}{2T}-\frac{1}{3}(\frac{b}{2T})^{3}+\frac{a}{2T}-\frac{1}{3}(\frac{a}{2T})^{3}}
{2T(a+b)}\right]-\frac{\widetilde{m}(\Lambda-k_{F})}{\pi^{2}}\\
=&4\pi\frac{1}{(2\pi)^{3}}\int^{\Lambda}_{k_{F}} k^{2}dk
\left[\frac{1}{4T^{2}}-\frac{1}{3}\frac{b^{3}+a^{3}}{(2T)^{4}}\frac{1}{a+b}
\right]-\frac{\widetilde{m}(\Lambda-k_{F})}{\pi^{2}}\\
=&4\pi\frac{1}{(2\pi)^{3}}\int^{\Lambda}_{k_{F}} k^{2}dk
\left[\frac{1}{4T^{2}}-\frac{1}{3}\left(\frac{1}{2T}\right)^{4}(a^{2}-ab+b^{2})
\right]-\frac{\widetilde{m}(\Lambda-k_{F})}{\pi^{2}}\\
=&\frac{1}{8\pi^{2}T^{2}}\frac{(\Lambda-k_{F})^{3}}{3}
-\mathcal{F}
-\frac{\widetilde{m}(\Lambda-k_{F})}{\pi^{2}}
\end{aligned}
\end{equation}
where
\begin{equation} 
\begin{aligned}
\mathcal{F}=&\frac{1}{6\pi^{2}}\frac{1}{(2T)^{4}}
             \int^{\Lambda}_{k_{F}}k^{2}(a^{2}-ab+b^{2})dk\\
=&\frac{1}{6\pi^{2}}\frac{1}{(2T)^{4}}
\frac{k^{3}}{420m_{\psi}^{2}m_{\phi}^{2}}
[
35m^{2}_{\psi}(4c^{2}m_{\phi}^{2}+2cm_{\phi}((p+q)^{2}-2dm_{\phi})\\
&+((p+q)^{2}-2dm_{\phi})^{2})
+21k^{2}m_{\psi}(2cm_{\phi}(m_{\psi}-2m_{\phi})+2dm_{\phi}(m_{\phi}-2m_{\psi})+(6m_{\psi}-m_{\phi})(p+q)^{2})\\
&-105km_{\psi}^{2}(p+q)(cm_{\phi}-2dm_{\phi}+(p+q)^{2})
+15k^{4}(m_{\psi}^{2}-m_{\phi}m_{\psi}+m_{\phi}^{2})\\
&-35k^{3}m_{\psi}(2m_{\psi}-m_{\phi})(p+q)
]\bigg|^{\Lambda}_{k_{F}}.
\end{aligned}
\end{equation}
where we define $c\equiv \mu_{\uparrow}$, $d\equiv i\omega+i\Omega+\mu_{\downarrow}$.

As can be seen, the ladder approximation (by summing over the ladder diagrams which correpond to the forward scattering) results in accurate results of the pair propagator and
self-energy as shown in the main text (at zero-temperature limit),
and it also agrees with the Quantum Monte-Carlo calculation as well as
the experimemtal results.
The single-channel $T$-matrix (which contains the pair propagator) introduces the tuneable $s$-wave scattering length to the 
manipulation of the behavior of a single impurity embedded to a fermi sea,
which describes the scattering between a pair of atoms with up and down spins, respectively,
and within the center of mass frame with energy $\varepsilon=\omega-(p+q)/2(m_{\psi}+m_{\phi})+\mu_{\uparrow}+\mu_{\downarrow}$.
At finite temperature and for the configuration that the number density of the bose impurity 
is much lower than the fermions (without the effect of three-atom loss (the Efimov.trimers)\cite{Fratini E,Massignan P,Pietil? V}), the pair 
propagator can also be written as
\begin{equation}
\begin{aligned}
\Pi(p+q,\omega+\Omega)
=&\frac{4\pi}{(2\pi)^{3}}\int^{\Lambda}_{k_{F}}k^{2}dk
    \frac{-1+N_{B}(\varepsilon_{p+q-k\downarrow})+N_{F}(\varepsilon_{k\uparrow}-\varepsilon_{q\uparrow})}{i\omega-s\varepsilon_{p+q-k\downarrow}-(s\varepsilon_{k\uparrow}-s'\varepsilon_{q\uparrow})}F_{ss'} 
-\frac{\widetilde{m}\Lambda}{\pi^{2}}.
\end{aligned}
\end{equation}
For pairing mechanism, this expression is definitely important,
e.g., for the pairing instability\cite{Cui X,Pietil? V,Adachi K,Pekker D} and the resonantly enhanced correlation,
and its real part and imaginary part are easy to obtained
by firstly replacing the imaginary frequencies in denominator with the 
analytical continuation and then using the Dirac identity (for retarded functions)
 $\lim_{\eta\rightarrow 0}\frac{1}{x\pm i\eta}=P(\frac{1}{x})\mp i\pi\delta(x)$.
We can see that the factor $F_{ss'}$ is in fact related to the angle between the wave vectors of
polaron (coherently dressed by the particle-hole excitations
of majority part) and the electron with momentum $k$.
And this term is unnecessary in the three (or two)-dimensional electron (or hole) gases,
it
is nonzero only when the eigenstates at different wave vectors have overlap
(corresponds to the two statistical functions in the numerator),
which for three-dimensional system (consider the longitudinal wave vector $k_{z}$) are
\begin{equation}
\begin{aligned}
\Psi_{+}=\begin{pmatrix}
e^{i\theta_{\parallel}}{\rm cos}\frac{\theta_{\bot}}{2}\\
{\rm sin}\frac{\theta_{\bot}}{2}
\end{pmatrix},\\
\Psi_{-}=\begin{pmatrix}
e^{i\theta_{\parallel}}{\rm sin}\frac{\theta_{\bot}}{2}\\
-{\rm cos}\frac{\theta_{\bot}}{2}
\end{pmatrix},
\end{aligned}
\end{equation}
where the indices $\pm$ denote the sign of band energy (i.e., the conduction band and valence band),
and $\theta_{\parallel}={\rm atan}k_{y}/k_{x}$,
    $\theta_{\bot}={\rm atan}\sqrt{k^{2}_{x}+k^{2}_{y}}/k_{z}$.
Then the overlap factor reads
$F_{ss'}=\frac{1\pm ({\rm cos}\theta_{\bot}{\rm cos}\theta'_{\bot}-{\rm sin}\theta_{\bot}{\rm sin}\theta'_{\bot}{\rm sin}\ b)}{2}$
where $\theta'_{\bot}={\rm atan}\sqrt{k'^{2}_{x}+k'^{2}_{y}}/k'_{z}$.
That are clearly different to the ones appear in two-dimensional system\cite{Culcer D}.
For the calculation in main text, we use the two-dimensional chiral factor $F_{ss'}=\frac{1\pm {\rm cos}\ b}{2}=\frac{1}{2}(1\pm \frac{k+q{\rm cos}\ a}{\sqrt{k^{2}+q^{2}+2kq{\rm cos}\ a}})$
due to the nature of weak-chirality of the system we discussed.
For another case of three-dimensional system, at long-wavelength limit ($k_{z}\rightarrow 0$) and with isotropic dispersion, 
such approximation is also applicable as shown in, e.g., Ref.\cite{Ahn S}.
When the vertex corrction is not taken into account,
for the case of inversed frequency $\Pi(p+q,-\omega-\Omega)$,
we can use the identity $\Pi(p+q,-\omega-\Omega)=\Pi^{*}(p+q,\omega+\Omega)$,
i.e., ${\rm Re}\ \Pi(p+q,-\omega-\Omega)={\rm Re} \Pi(p+q,\omega+\Omega)$,
${\rm Im}\ \Pi(p+q,-\omega-\Omega)=-{\rm Im} \Pi(p+q,\omega+\Omega)$
Further, when the chirality (from the Weyl system) appears, the causality relations are studied in Ref.\cite{Zhou J}.

Then the self-energy at finite temperature can be obtained as
\begin{equation} 
\begin{aligned}
\Sigma(p,\omega)=\frac{T}{V}\int^{k_{F}}_{0}\frac{d^{3}q}{(2\pi)^{3}}\sum_{n}
T(p+q,\omega+\Omega)G^{0}_{\psi}(q,\Omega),
\end{aligned}
\end{equation}
where $G^{0}_{\psi}(q,\Omega)$ can also be replaced by $G_{\psi}(q,\Omega)$
which contains the self-energy term when consider the self-energy effect as done in Ref.\cite{Enss T} with strong scattering strength.
In addition, we discuss the case when consider the ladder vertex correction,
where summation over Matsubara frequency can be done by using the method of coutour integral (in optical limit)\cite{Mahan G D},
\begin{equation} 
\begin{aligned}
\frac{T}{V}\sum_{n'}G^{\psi}(\nu)G^{\phi}(\omega+\Omega-\nu)\Gamma(\nu,\omega+\Omega-\nu)=-\oint_{\mathcal{C}}\frac{dz}{2\pi i}G^{\psi}(z)G^{\phi}(\omega+\Omega-z)\Gamma(z,\omega+\Omega-z),
\end{aligned}
\end{equation}
where $\Gamma(\nu,\omega+\Omega-\nu)$ denotes the vertex function.

At finite (low) temperature where the $s$-wave scattering is still dominating,
the low-energy excitations induced by quantum fluctuation has a more significant effect on the properties of polaron 
compared to the thermal excitations
especially for the case of small-chemical potential,
like the particle-hole parts (especially at low dimension\cite{Koschorreck M,Endres M}) or the phonon-like (Fr{\o}hlich type) excitations.

\end{large}
\renewcommand\refname{References}

\clearpage

Fig.1
\begin{figure}[!ht]
   \centering
 \centering
   \begin{center}
     \includegraphics*[width=0.6\linewidth]{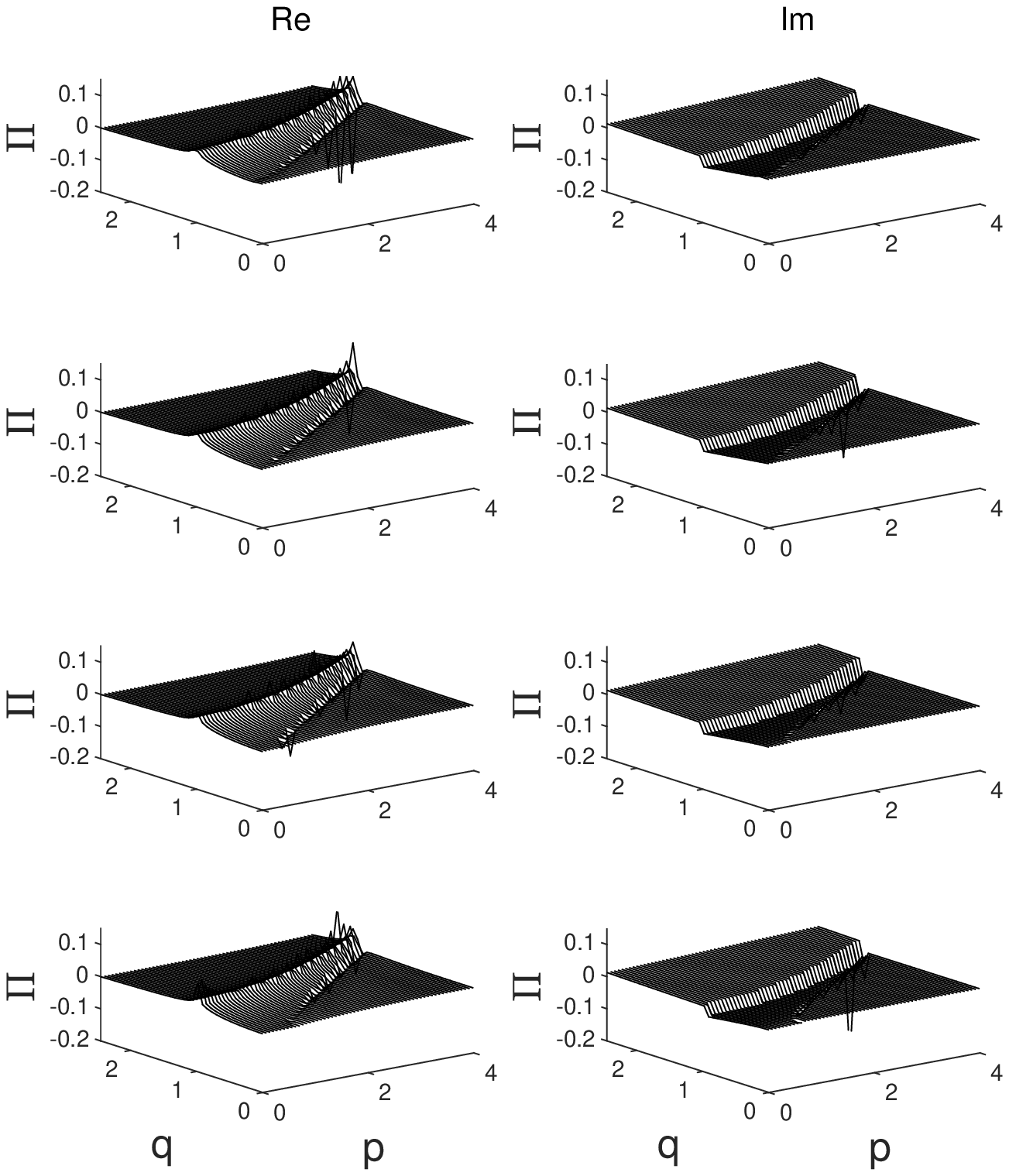}
\caption{Real part (left) and imaginary part (right) of the pair propagator at non-chiral case
as a function of the impurity momentum $p$ and majority momentum $q$.
The rows from top to bottom correspond to the Bosonic frequency (impurity) $\omega=-1,\ 0,\ 1,\ 2$, respectively.
The momentum cutoff $\Lambda$ is setted as 1 and the chemical potential is zero.
The vertical axis is in unit of $\frac{1}{2\pi}$.
}
   \end{center}
\end{figure}
\clearpage
Fig.2
\begin{figure}[!ht]
   \centering
 \centering
   \begin{center}
     \includegraphics*[width=0.6\linewidth]{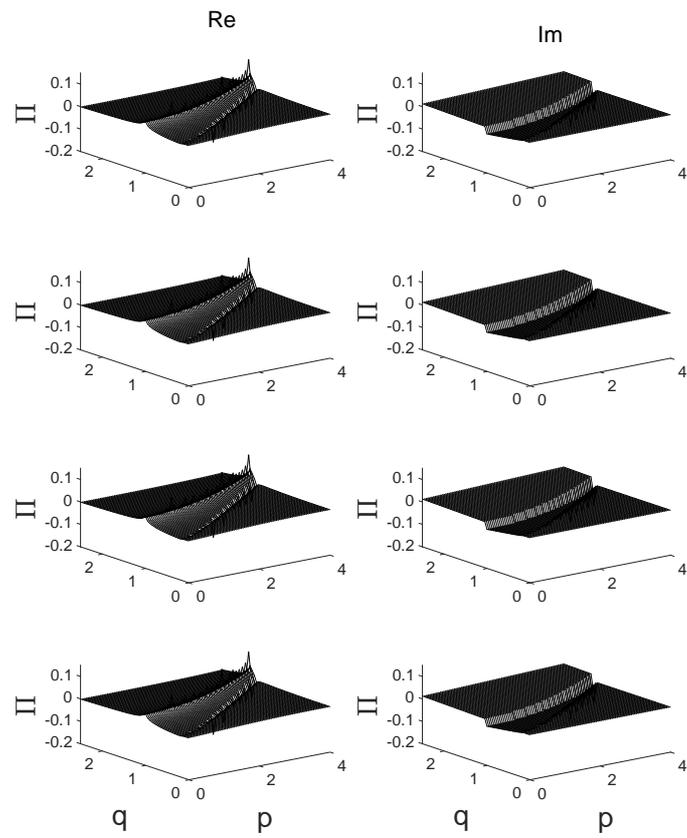}
\caption{The same as Fig.1 but for chemical potential $\mu_{\uparrow}=0.5$.
}
   \end{center}
\end{figure}
\clearpage
Fig.3
\begin{figure}[!ht]
   \centering
 \centering
   \begin{center}
     \includegraphics*[width=0.9\linewidth]{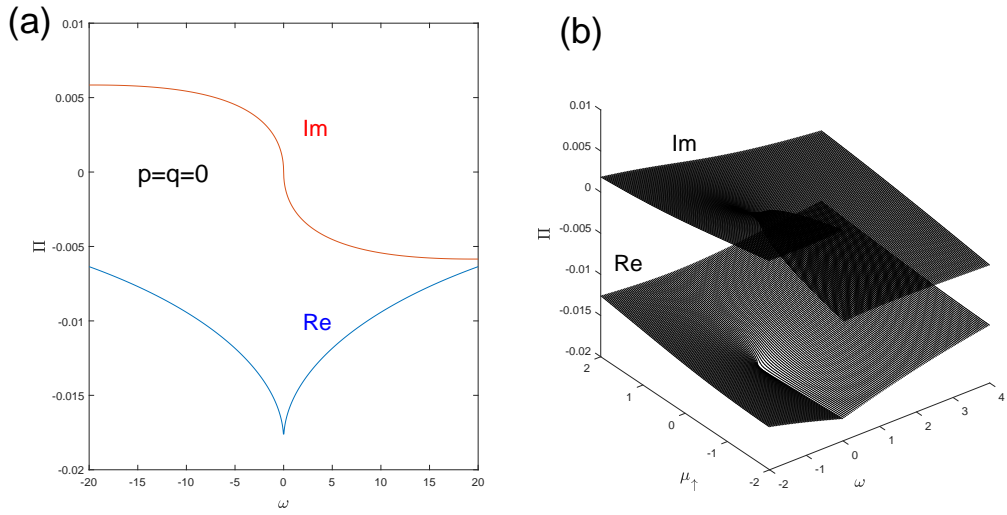}
\caption{(a) Real part and imaginary part of the (non-chiral) pair propagator at $p=q=0$ with zero chemical potential
as a function of the Bosonic frequency (impurity) $\omega$.
(b) The pair propagator at $p=q=0$ with nonzero chemical potential.
}
   \end{center}
\end{figure}

\clearpage
Fig.4
\begin{figure}[!ht]
   \centering
 \centering
   \begin{center}
     \includegraphics*[width=0.6\linewidth]{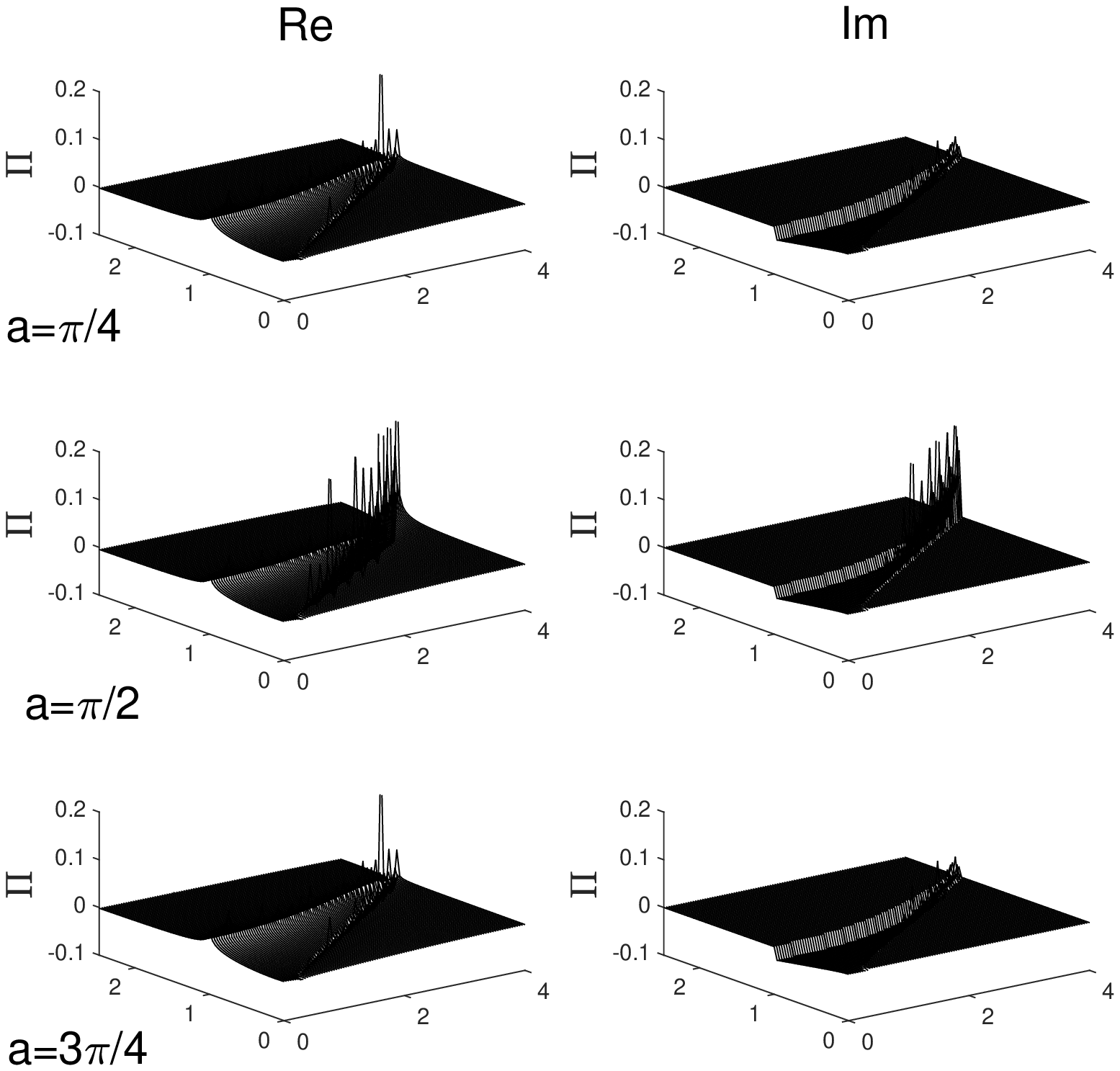}
\caption{Intraband part of the pair propagator where the chiral factor is taken into account.}
   \end{center}
\end{figure}

Fig.5
\begin{figure}[!ht]
   \centering
 \centering
   \begin{center}
     \includegraphics*[width=0.6\linewidth]{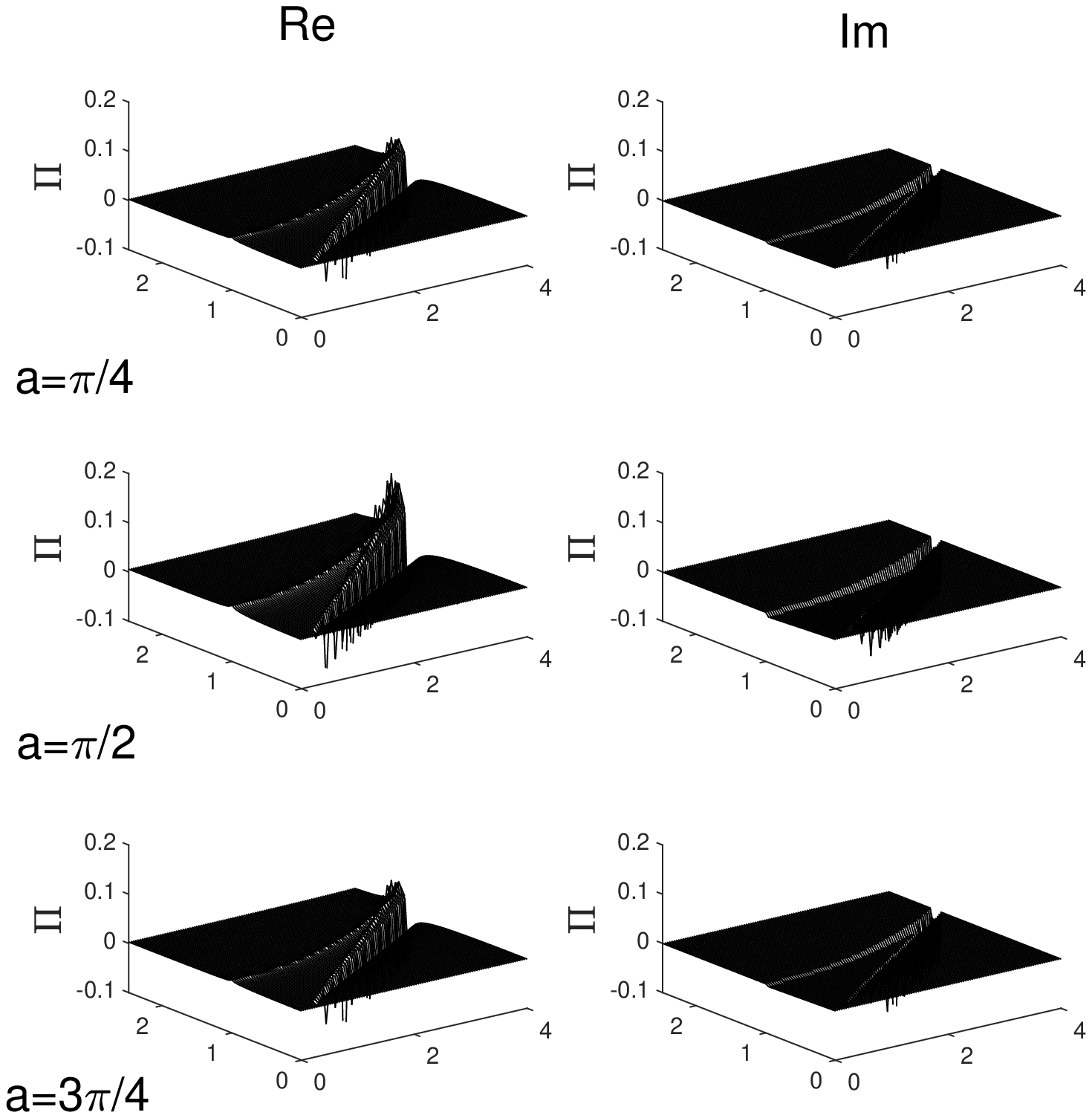}
\caption{Interband part of the pair propagator where the chiral factor is taken into account.}
   \end{center}
\end{figure}
\clearpage

Fig.6
\begin{figure}[!ht]
   \centering
 \centering
   \begin{center}
     \includegraphics*[width=0.6\linewidth]{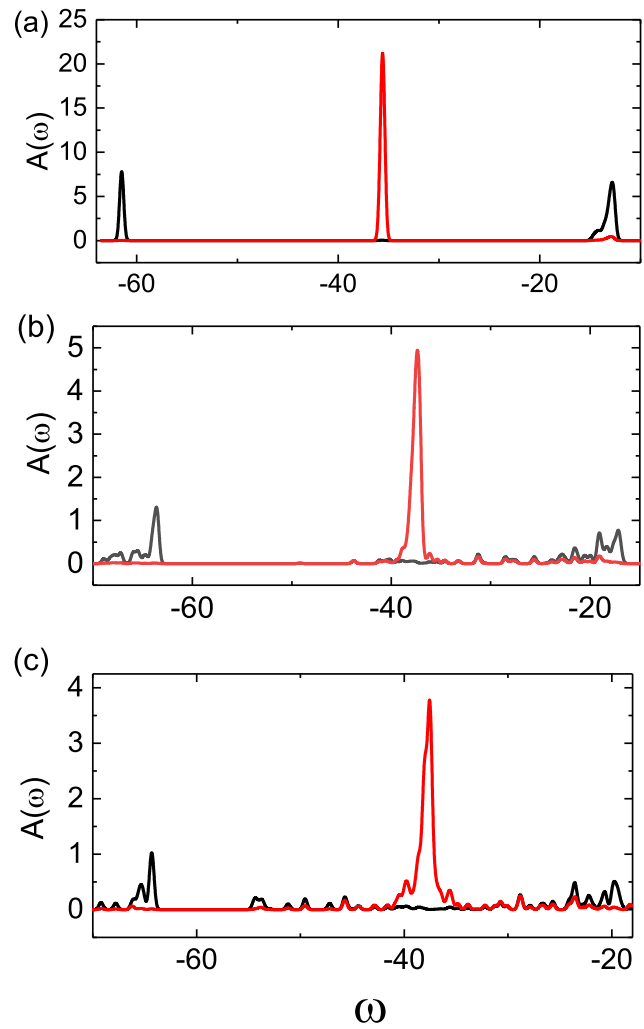}
\caption{
Spectral function (the local density of states) of the hole states at zero momentum (p=0) in the Fig.4 
with (b,c) and without (a) the Hubbard interaction.
The strength of the Hubbard interaction is (c) is setted larger than that in (b).
While the density of states can be obtained by integration over the momentum space.
The many-electron effect is contained in this figure as can be seen from the broaden peaks.
The red-line and the black-line corresponds to the contributions from the $p$-orbit electrons and $s$-orbit electrons, respectively.}
   \end{center}
\end{figure}
\clearpage
Fig.7
\begin{figure}[!ht]
   \centering
 \centering
   \begin{center}
     \includegraphics*[width=1\linewidth]{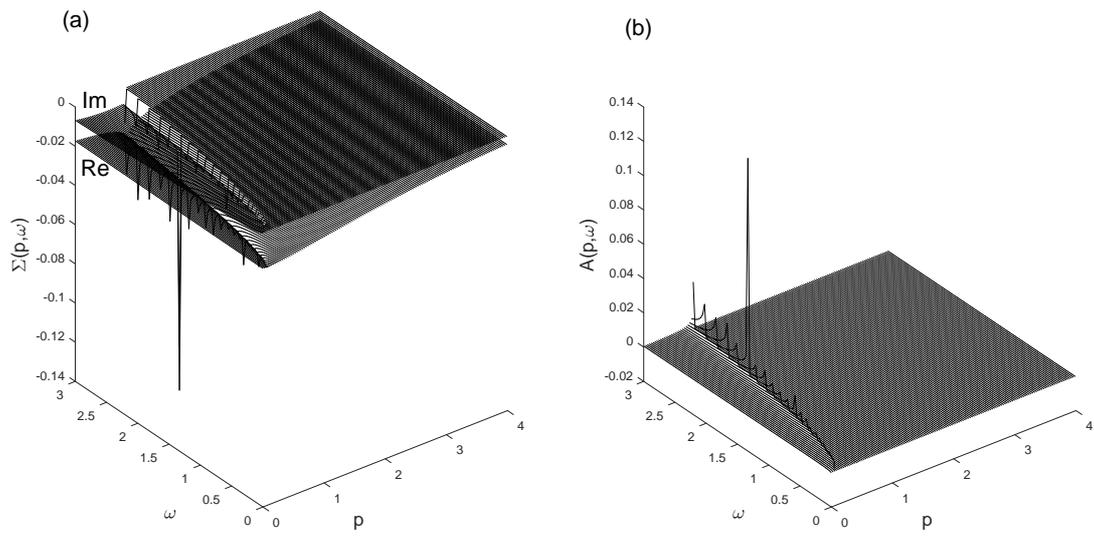}
\caption{Polaron self-energy (a) and 
real part of the spectral function (b) in the presence of quantum many-body effect.}
   \end{center}
\end{figure}
\clearpage
Fig.8
\begin{figure}[!ht]
   \centering
 \centering
   \begin{center}
     \includegraphics*[width=1\linewidth]{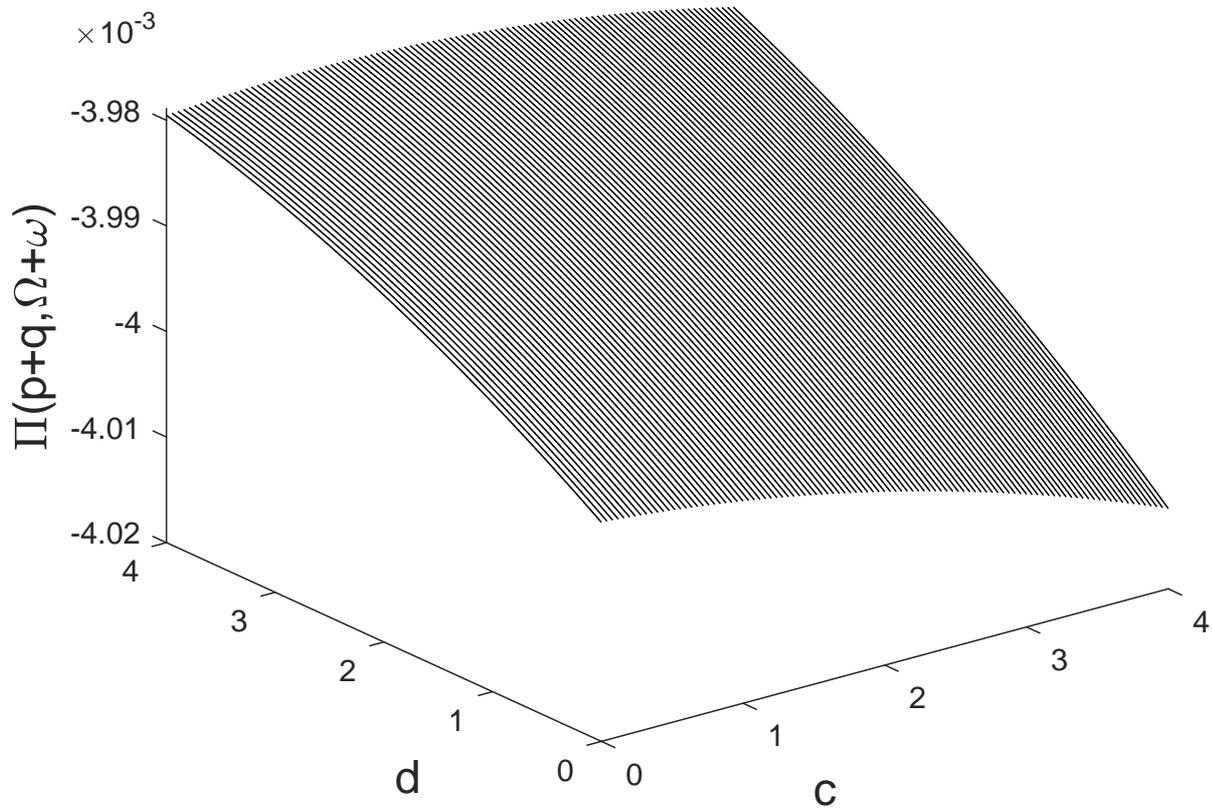}
\caption{Pair propagator at finite temperature as a function of the parameters $c$ and $d$ (see Appendix. B).
We set
$T=5$ K, $\Lambda=1$, $k_{F}=0.2$, $m_{\phi}=m_{\psi}=0.1$ here.
}
   \end{center}
\end{figure}

\end{document}